\documentclass[twoside,a4paper]{article}
\usepackage{amsmath,graphicx,amssymb,fancyhdr,amsthm,enumerate,textcomp} 
\usepackage[usenames]{color}
\pdfoutput=1
\newtheorem{thm}{Theorem}[section] 
\newtheorem{cor}[thm]{Corollary} 
 
\newtheorem{prop}[thm]{Proposition} 
 
\theoremstyle{definition} 
\newtheorem{defn}[thm]{Definition}
  
\theoremstyle{remark}  
  
\def\beq{\begin{eqnarray}}  
\def\eeq{\end{eqnarray}}  
\def\bsp{\begin{split}}  
\def\esp{\end{split}}

\def\d{\mathrm{d}}

\def\b{\tilde{b}}

\def\VC{V^{\mathbb{C}}}
\def\GC{G^{\mathbb{C}}}

\newcommand{\mf}[1]{{\mathfrak #1}}   
   
\newcommand{\mb}[1]{{\mathbb #1}}   
   
\newcommand{\mbold}[1]{\mbox{\boldmath{\ensuremath{#1}}}}

\begin{document}   
   
\title{\Large\textbf{A spacetime not characterised by its invariants is of aligned type II}}  
\author{{\large\textbf{Sigbj\o rn Hervik  }    }
 \vspace{0.3cm} \\    
Faculty of Science and Technology,\\    
 University of Stavanger,\\  N-4036 Stavanger, Norway    \\
\vspace{0.3cm} \\     
\texttt{sigbjorn.hervik@uis.no} }    
\date{\today}    
\maketitle  
\pagestyle{fancy}  
\fancyhead{} 
\fancyhead[EC]{S. Hervik}  
\fancyhead[EL,OR]{\thepage}  
\fancyhead[OC]{The alignment theorem}  
\fancyfoot{} 
  
\begin{abstract} 
By using invariant theory we show that a (higher-dimensional) Lorentz\-ian metric that is not characterised by its invariants must be of aligned type II; i.e., there exists a frame such that all the curvature tensors are simultaneously of type II. This implies, using the boost-weight decomposition, that for such a metric there exists a frame such that all positive boost-weight components are zero. Indeed, we show a more general result, namely that  any set of tensors which is not characterised by its invariants, must be of aligned type II. This result enables us to prove a number of related results, among them the algebraic VSI conjecture. 
\end{abstract}
\section{Introduction}
Recently there has been an interest in the relation between metrics and their \emph{polynomial curvature invariants}\footnote{Here, we will be interested in the polynomial curvature invariants and so in what follows 'invariants' is to be understood as '\emph{polynomial invariants}'. These can always be considered as full contractions of the curvature tensors \cite{GW}.}. One of the first class of metrics that was investigated in this regard was the class of metrics  having all vanishing curvature invariants (VSI spacetimes). The 4 dimensional case was considered in \cite{4DVSI} and it was proven that all VSI spacetimes were of Kundt type and they had curvature tensors of aligned type III or simpler. The higher-dimensional case was considered in \cite{Higher} and it was indicated that the same was the case: the VSI spacetimes are of Kundt type and have aligned type III curvature tensors. However, a thorn in the proof is the apparent lack of a proof of Lemma 10. 

For the spacetimes having constant scalar invariants (CSI spacetimes) the degenerate Kundt again were of importance \cite{CSI,3DCSI,4DCSI}. Indeed, in 4 dimensions, all spacetimes having a degenerate curvature structure in the sense that the metric is not characterised by the curvature invariants,  are the degenerate Kundt spacetimes \cite{inv}. In terms of the algebraic classification, all metrics not being characterised by its invariants are of aligned type II to all orders. 

We are thus led to an important question. Does any Lorentzian metric in higher dimensions which is not characterised by its polynomial curvature invariants have curvature tensors of type II to all orders? Moreover, are they aligned, i.e., are they all of type II in the same frame? In this paper we will use invariant theory and explore this question. Indeed, our result is that the answer to this question is yes: a metric not characterised by its invariants has curvature tensors of type II, and they are all aligned. We will refer to this result as the \emph{alignment theorem}. 

In order to prove this theorem we will take the ideas in \cite{epsilon2} one step further and use real invariant theory and group theory, combined with standard real analysis. This is a new approach to this problem, but it is highly successful; indeed, some of the corollaries of this theorem  is a proof of the algebraic VSI conjecture \cite{Higher} (thereby also Lemma 10 therein) along with proofs of several conjectures in \cite{CSI,invHigher}. 
\subsection{Boost weight decomposition}
An important tool in the classification of spacetimes in higher dimensions is the boost weight decomposition \cite{class,Milson} which we will review in brief.

Given a covariant tensor $T$ with respect to a null
frame, $\{\ell,n,m^i\}$, the effect of a boost $\ell \mapsto e^{\lambda}\ell$, $n
\mapsto e^{-\lambda}n$ allows $T$ to be decomposed according to
its boost weight components
\begin{equation}
T=\sum_b (T)_{b} \label{bwdecomp}
\end{equation}
where $(T)_{b}$ denotes the boost weight $b$ components (with respect to the abovementioned boost) of $T$. An
algebraic classification of tensors $T$ has been developed
\cite{class,Milson} which is based on the existence of certain
normal forms of (\ref{bwdecomp}) through   successive application
of null rotations and spin-boost.  In the special case where $T$
is the Weyl tensor in four dimensions, this classification reduces
to the well-known Petrov classification. However, the boost weight
decomposition can be used in the classification of any tensor $T$
in arbitrary dimensions.  As an application, a Riemann tensor of type $G$ has the following decomposition
\begin{equation}
R=(R)_{+2}+(R)_{+1}+(R)_{0}+(R)_{-1}+(R)_{-2}
\end{equation}
in every null frame.  A Riemann tensor is algebraically special if there exists a frame in which certain boost weight components can be transformed to zero, these are summarized in Table \ref{riemtypes}.

In general we can define the following algebraic special cases which we will be useful for us: 
\begin{defn}
A tensor, $T$, is of 
\begin{itemize} 
\item type II if there exists a frame such that all the positive boost-weight components are zero, $(T)_{b>0}=0$; 
\item type D if there exists a frame such that $T$ has only boost-weight 0 components,  $T=(T)_{0}$; 
\item type III if there exists a frame such $T$ has only negative boost weights,  $(T)_{b\geq 0}=0$.
\end{itemize}
\end{defn}
\noindent If the tensor is of neither of these cases the tensor is of type I/G, or more general \cite{class,Milson}. 

This implies that a tensor, $T$, of type II has only non-positive boost  weight components: 
\[ T=\sum_{b \leq 0}(T)_b, \qquad (\text{type II}). \] 
For two tensors, $T$ and $S$, both of type II, then these have, separately, the above boost-weight decomposition, however, they may be with respect to \emph{different frames}: 
\[ T=\sum_{b \leq 0}(T)_b, \quad S=\sum_{\tilde{b} \leq 0}(S)_{\b}.\] 
If the tensors are of type II \emph{in the same frame}, then we say that the tensors $T$ and $S$ are of \emph{aligned type II}. 

A useful discrete symmetry is the following (orientation-reversing) Lorentz transformation:
\beq
\ell\leftrightarrow n, 
\label{dsym}\eeq
which interchanges the boost weights, $(T)_b\rightleftharpoons (T)_{-b}$. 

\begin{table}[h]
\begin{center}
\small{
\begin{tabular}{c|c}
\hline
 & \\
{\bf Riemann type} & {\bf Conditions}  \\
 & \\
\hline
G & --- \\
I & $(R)_{+2}=0$ \\
II & $(R)_{+2}=(R)_{+1}=0$ \\
III & $(R)_{+2}=(R)_{+1}=(R)_{0}=0$ \\
N & $(R)_{+2}=(R)_{+1}=(R)_{0}=(R)_{-1}=0$ \\
D & $(R)_{+2}=(R)_{+1}=(R)_{-1}=(R)_{-2}=0$ \\
O & all vanish (Minkowski space) \\
\hline
\end{tabular}
}
\caption{The relation between Riemann types and the vanishing of boost weight components.  For example, $(R)_{+2}$ corresponds to the frame components $R_{1i1j}$.}\label{riemtypes}
\end{center}
\end{table}

\subsection{Invariant theory}
Let us review some of the results from invariant theory, specifically from \cite{procesi,eberlein}. 

The idea is to consider a group $G$ acting on a vector space $V$. In our case we will be consider a real $G$ and a real vector space $V$. However, it is adventageous to review the complex case with a complex group $\GC$ acting on a complex vector space $\VC$.  Then for a vector $X\in \VC$ we can define the \emph{orbit} of $X$ under the action of $\GC$ as follows: 
\[ \mathcal{O}_{\mathbb{C}}(X)\equiv \{  g(X)\in\VC ~\big{|}~g\in \GC \}\subset \VC\]

Then (\cite{procesi}, p555-6): 
\begin{thm}
If $\GC$ is a linearly reductive group acting on an affine variety $\VC$, then the ring of invariants is finitely generated. Moveover, the quotient $\VC/ \GC$ parameterises the closed orbits of the $\GC$-action on $\VC$ and the invariants separate closed orbits.   
\end{thm}

Here the term \emph{closed} refers to \emph{topologically closed} with respect to the standard vector space topology and henceforth, closed will mean topologically closed. 
This implies that given two distinct closed orbits $A_1$ and $A_2$, then there is an invariant with value $1$ on $A_1$ and $0$ on $A_2$. This enables us to define the set of orbits: 
\beq
\mf{C}_{\mathbb{C}}=\{\mathcal{O}_{\mathbb{C}}(X)\subset \VC ~\big{|}~ \mathcal{O}_{\mathbb{C}}(X) \text{ closed.}  \}
\eeq
Based on the above theorem we can thus say that the invariants separate elements of $\mf{C}_{\mb{C}}$ and hence we will say that an element of  $\mf{C}_{\mb{C}}$ is \emph{characterised by its invariants}. 

In our case we will consider the real case where we have the Lorentz group, $O(1,n-1)$ which is a real semisimple group. For real semisimple groups acting on a real vector space we do not have the same uniqueness result as for the complex case \cite{eberlein}. However, by complexification, $[G]^{\mathbb{C}}=\GC$ we have $[O(1,n-1)]^{\mathbb{C}}=O(n,\mathbb{C})$, and by complexification of the real vectorspace $V$ we get $\VC\cong V+iV$. The complexification thus lends itself to the above theorem, and consequently we will define \emph{characterised by its invariants} as follows. For a tensor, $T$, a rotation of a frame naturally defines a group action on the \emph{components} of the tensor. Then:
\begin{defn}
Consider a (real) tensor, $T\in V$, or a direct sum of tensors, then if the orbit of the components of $T$ under the complexified Lorentz group $G^{\mathbb{C}}$ is an element of $\mf{C}_{\mb{C}}$, i.e., $\mathcal{O}_{\mathbb{C}}(T) \in \mf{C}_{\mathbb{C}}$, then we will say that $T$ is \emph{characterised by its invariants}. 
\end{defn}
\noindent As the invariants parameterise the set $\mf{C}_{\mb{C}}$ and since the group action defines an equivalence relation between elements in the same orbit this definition makes sense.

Let us similarly define the real orbits: 
\[\mathcal{O}(X)\equiv \{ g(X)\in V ~\big{|}~g\in G \}\subset V\]
How do these results translate to the real case? The real orbit $\mathcal{O}(T)$, is a real section of the complex orbit  $\mathcal{O}_{\mathbb{C}}(T)$. However, there might be more than one such real section having the same complex orbit. Using the results of \cite{eberlein}, these real closed orbits are disjoint, moreover: 
\begin{thm}
$\mathcal{O}(T)$ is closed in $V$ $\Leftrightarrow$ $\mathcal{O}_{\mathbb{C}}(T)$ is closed in $\VC$.
\end{thm}
Thus the question of whether $T$ is characterised by its invariants is equivalent to whether $\mathcal{O}(T)$ is closed in $V$. Thus we can define similarly: 
\beq
\mf{C}=\{\mathcal{O}(X)\subset V ~\big{|}~ \mathcal{O}(X) \text{ closed.}  \},
\eeq
hence, we have that $T$ is characterised by its invariants iff $\mathcal{O}(T)\in\mf{C}$.

 However, as pointed out, there might be other closed real orbits  $\mathcal{O}(\tilde{T})$ having the same invariants as  $\mathcal{O}({T})$ (in line with the comments in \cite{inv,operator}). An example of this is the pair of metrics:
\beq
\d s^2_1&=& -\d t^2+\frac{1}{x^2}\left(\d x^2+\d y^2+\d z^2\right),\nonumber \\
\d s^2_2&=& \d \tau^2+\frac{1}{x^2}\left(\d x^2+\d y^2-\d \zeta^2\right),
\eeq
The curvature tensors\footnote{Both of these metrics are symmetric and conformally flat, so the only non-zero curvature tensor is the Ricci tensor.}  of these metrics lie in separate orbits $\mathcal{O}(T)$, but in the same complex orbit $\mathcal{O}_{\mathbb{C}}(T)$.  

Note that the action of a $g\in G$ on the components of a tensor $T$, $g(T)$, can be interpreted as expressing the components in a different frame. If we consider a $h\in G$, we can write, 
\[ hg(T)=hgh^{-1} h(T).\] 
The adjoint map $G\mapsto h G h^{-1}$ is an isomorphism of Lie groups, implying that if $g$ is, for example, a boost, then so will $hgh^{-1}$ be but with respect to a different frame. 

\subsection{The orbit $\mathcal{O}(X)$ as a dynamical system} 
Another useful view of the orbits $\mathcal{O}(X)$ is as a dynamical system. Consider the one-parameter groups $g_{\tau}\equiv e^{\tau {\mathcal X}}$, where ${\mathcal X}$ is an element of the Lie algebra of $O(1,n-1)$; i.e., ${\mathcal X}\in \mathfrak{o}(1,n-1)$. Then $g_{\tau}$ is generated by ${\mathcal X}$ and is in the connected component of $O(1,n-1)$. 

We can now view the orbit $\mathcal{O}(X)$ as a dynamical system with a flow generated by ${\mathcal X}$ as follows. Consider a $Y_0\in \mathcal{O}(X)$. Then we can define the flow $Y_{\tau}$ as $Y_{\tau}=g_{\tau}(Y_0)$. The $Y_{\tau}=Y(\tau)$ fulfills the linear differential equation: 
\beq 
\frac{ d Y}{d\tau}={\mathcal X}(Y), \quad Y(0)=Y_0. 
\label{dynsys}\eeq
This differential equation, for each ${\mathcal X}\in\mathfrak{o}(1,n-1)$ defines a flow on $\mathcal{O}(X)$. The unique solutions to this equation are the one-parameter orbits $Y_{\tau}=g_{\tau}(Y_0)$. 

As an example, if ${\mathcal X}$ generates a boost, $b$, then the decomposition  $T_0=\sum_b(T_0)_b$ is an \emph{eigenvector decomposition} with respect to this boost. Each eigenvector $(T_0)_b$ has with eigenvalue $b$, i.e., $ {\mathcal X}\left((T)_b\right)=b(T)_b$; hence, by integration $Y_\tau=\sum_b e^{b\tau}(T_0)_b$. The boost weight decomposition is thus an eigenvalue decomposition with respect to the differential equation (\ref{dynsys}).

These dynamical systems are useful for us because there always exists a flow such that it has the point on the boundary as a limit. In particular, we have the following result (see, e.g., \cite{eberlein}, Proposition 1.6):
\begin{prop}
Let $X\in V$ and assume that $\mathcal{O}(X)$ is not closed. Then there exists an ${\mathcal X}\in \mathfrak{o}(1,n-1)$ and an $X_0\in V$ such that $e^{\tau{\mathcal X}}(X)\rightarrow X_0$ as $\tau\rightarrow\infty$.  Furthermore, the orbit  $\mathcal{O}(X_0)$ is closed.
\label{flowprop}\end{prop}

Indeed, this Lie algebra element ${\mathcal X}$ can be chosen among the elements conjugate to the null-rotations and boosts (see later and \cite{eberlein}). 

\section{The Alignment theorem}
Consider a set of tensors $R^{(i)}$, $i=1,...,N$. We will express these tensors in terms of a basis  of orthonomal (or null) vectors $\omega=\{ {\mbold e}_1,...,{\mbold e}_n\}$ for an $n$-dimensional Lorentzian manifold. 
Define the vector consisting of the components: $X=[ R^{(1)}_{a_1a_2...a_{k_1}}, R^{(2)}_{a_1a_2...a_{k_2}},...,R^{(N)}_{a_1a_2...a_{k_N}}]\in \mathbb{R}^m$, for some appropriate $m$. 

Our main result is the following. 
\begin{thm}[The alignment theorem]
Assume that the set of tensors  $R^{(i)}$, $i=1,...,N$ are not characterised by their invariants. Then there exists a null-frame such that all tensors have, in the same frame, the boost weight decomposition: 
\[ X=(X)_0+\sum_{b<0}(X)_{b}, \] 
i.e., all positive boost weight components are zero. In particular, this means that all the tensors  $R^{(i)}$ are of type II and are \emph{aligned}. 
\end{thm} 

\subsection{Proof of the alignment theorem}
The proof of this theorem goes as follows. 

The vector $X=[ R^{(1)}_{a_1a_2...a_{k_1}}, R^{(2)}_{a_1a_2...a_{k_2}},...,R^{(N)}_{a_1a_2...a_{k_N}}]\in \mathbb{R}^m$ can be considered to generate an orbit under the Lorentz group $O(1,n-1)$ (which is semisimple). The action corresponds to a frame rotation and explicitly, if we consider the matrix $g=(M^a_{~b})\in O(1,n-1)$, acting as a frame rotation $g\omega=\{ M^a_{~1}{\mbold e}_a,..., M^a_{~n}{\mbold e}_a\}$, the frame rotation induces an action on $X$ through the tensor structure of the components:
\[ g(X)=\left[ M^{b_1}_{~a_1}...M^{b_{k_1}}_{~a_{k_1}}R^{(1)}_{b_1...b_{k_1}},M^{b_1}_{~a_1}...M^{b_{k_2}}_{~a_{k_2}}R^{(2)}_{b_1...b_{k_2}},...,M^{b_1}_{~a_1}...M^{b_{k_N}}_{~a_{k_N}}R^{(N)}_{b_1...b_{k_N}}\right] \]
The orbit $\mathcal{O}(X)$ is now defined by:
\[\mathcal{O}(X)\equiv \{ g(X)\in \mathbb{R}^m ~\big{|}~g\in O(1,n-1) \}\subset \mathbb{R}^m.\]

Since the set is not characterised by its invariants, the orbit $\mathcal{O}(X)$ under the Lorentz group is not closed. Therefore,  the subset consisting of all points in the closure, $\overline{\mathcal{O}(X)}$, but not in the orbit, is non-empty; i.e., $\overline{\mathcal{O}(X)}-{\mathcal{O}(X)}\neq \emptyset$. Choose an element $p\in \overline{\mathcal{O}(X)}-{\mathcal{O}(X)}$. Since $p$ is in the closure there exists a sequence $p_n\in \mathcal{O}(X)$ which converges to $p$. This implies further that there exists a sequence $g_n\in G$ such that $g_n(X)=p_n$. 

First we will show that there is a subsequence $\tilde{g}_n$ of $g_n$ for which $||\tilde{g}_n||\rightarrow \infty$ (using the standard norm in the space of matrices). Assume that no such subsequence exists. Then this implies that $g_n$ is bounded. Consequently, there is a compact set, $U\subset G$, for which $g_n\in U$. However, since the group action: $f: G\mapsto \mathbb{R}^m$, given by $f(g)=g(X)$ is continuous, then the set $f(U)\subset \mathcal{O}(X)$ must be compact. However, since this implies that $p\in f(U) \subset \mathcal{O}(X)$ this is a contradiction. 

We can therefore assume that $||{g}_n||\rightarrow \infty$ by taking an appropriate subsequence. Let us use the Iwawsava decomposition $G=ANK$, where $A$ is a boost, $N$ is a null-rotation, and $K$ is the largest compact subgroup. For the Lorentz group $O(1,n-1)$, $\dim(A)=1$, $\dim(N)=(n-2)$ and $K=O(n-1)\times D$, where $D$ is a discrete group. Furthermore, the homeomorphism $G=ANK\cong \mathbb{R}\times\mathbb{R}^{n-2}\times K$, enables us to parameterise a group element $g\in G$ as a triple: $(\lambda,z,k)\in\mathbb{R}\times\mathbb{R}^{n-2}\times K$. Explicitly, we can choose $k\in K$, while we choose $\lambda$ and $z^i$ as vectors in the Lie algebras of $A$ and $N$, respectively\footnote{Recall that for a connected and simply connected Lie group the exponential map is a smooth (in fact analytic) diffeomorphism mapping the Lie algebra onto the Lie group \cite{Chevalley}.}. The sequence $g_n$ thus can be parameterised as $(\lambda_n,z_n,k_n)$. Consider first the sequence $k_n\in K$. Since $K$ is compact, there is a subsequence $k_{n_i}$ which converges to a $k\in K$: $k_{n_i}\rightarrow k$. By choosing the corresponding $\lambda_{n_i}$, and $z_{n_i}$ we get a subsequence $g_{n_i}$ so that $g_{n_i}(X)\rightarrow p$. We define now $\tilde{X}=k(X)$ which is also in $\mathcal{O}(X)=\mathcal{O}(\tilde{X})$. Using the previous subsequence $g_{n_i}$, we can now define a $\tilde{g}_{n_i}$ by the points $(\lambda_{n_i},z_{n_i}, 1_K)$, where $1_K$ is the unit in $K$. Then the sequence $\tilde{g}_{n_i}(\tilde{X})$ is a sequence in $\mathcal{O}(X)$ converging to $p$. 

By choosing the subsequence, $\tilde{g}_{n_i}(\tilde{X})$ if necessary, we can assume that we have a sequence $g_n(X)\rightarrow p$, for which can be parameterised as $(\lambda_n,z_n,1_K)$, and that $||g_n||\rightarrow \infty$. 

If either $\lambda_n$ or $z_n$ are bounded, we can similarly consider a compact region in $\mathbb{R}$ or $\mathbb{R}^{n-2}$ respectively. Thus we can choose subsequences which converge to $\lambda_{n_i}\rightarrow \lambda$ or $z_{n_i}\rightarrow z$. As above, we can consider the sequence $(\lambda_{n_i},0_{N},1_{K})$ acting on $\tilde{X}=N(X)$, (which converges to $p$) or  $(0,z_{n_i},1_{K})$, which converges to $B_{\lambda}^{-1}(p)$. We note that $B_{\lambda}^{-1}(p)\in \overline{\mathcal{O}(X)}-{\mathcal{O}(X)}$. In these cases we can thus assume, after a redefinition and a reparameterisation, that the sequence $g_n(X)\rightarrow p$, for which can be parameterised as $(\lambda_n,z_n,1_K)$, and that $||g_n||\rightarrow \infty$. 

Therefore, we have three possibilities left to consider: 
\begin{itemize}
\item[1)]{} $||\lambda_n||\rightarrow \infty$, $z_n=0$. 
\item[2)]{} $||z_n||\rightarrow \infty$, $\lambda_n=0$. 
\item[3)]{} Both $||\lambda_n||, ~||z_n||\rightarrow \infty.$
\end{itemize}
Let us consider these in turn using the boost weight decomposition (with respect to $B_{\lambda}$) of an arbitrary tensor $T$:
\[ T=\sum_{b=-N}^N(T)_b.\] 

\noindent \underline{1) $||\lambda_n||\rightarrow \infty$:}
Using the boost-weight decomposition we have the action of the boost of a tensor $T$: 
\[ B_\lambda[T]=...+e^{2\lambda}(T)_{+2}+e^{\lambda}(T)_{+1}+(T)_{0}+e^{-\lambda}(T)_{-1}+e^{-2\lambda}(T)_{-2}+...\] 
Since the point $p$ is finite, it implies that all the components must remain finite as $||\lambda_n||\rightarrow \infty$. This requires that either $\lambda\rightarrow \infty$ and $(T)_{b>0}=0$, or $\lambda\rightarrow -\infty$ and $(T)_{b<0}=0$. 

\noindent \underline{2) $||z_n||\rightarrow \infty$: } Using the boost-weight decomposition we can express the components $(T)_b$ as the action  $(T)_b=(N[z^i,T])_b$ in terms of the pre-transformed components $(T)_{\b}$ as: 
\[ (N[z^i,T])_b=(T)_{\b}+z^iP_i[ (T)_{\b+1}]+z^{i}z^jP_{ij}[(T)_{\b+2}]+...+z^{i_1}...z^{i_{N-\b}}P_{i_1...i_{N-\b}}[(T)_{\tilde{N}}],\] 
where $z^i$ are the components of $z$, and $P_{ij...k}[(T)_{\b}]$ is some linear combination of the boost-weight $b$ components of $T$. These are therefore polynomials in $z^i$ which are in general unbounded. As the unit vector $z^i/||z||$ takes values on the unit sphere (which is compact), we first choose a subsequence such that $ z_n^i/||z_n||$ converges to a vector $w^i$ on the unit sphere. Since $||z||\rightarrow \infty$, and the limit $p$ should be finite, we need that $z_n^i/||z_n||\rightarrow w^i$, where $\mathbb{R}w^i$ is an \emph{isotropy} of $T$; i.e., $(N[\mathbb{R}w^i,T])_b=(T)_{\b}$. We can therefore consider the quotient $N/N[w^i]$  (since the action of the isotropy group is trivial) and reduce the space of null-rotations to $\mathbb{R}^{n-3}$. Then considering the quotient, we get a sequence, $\hat{z}_n\in\mathbb{R}^{n-3}$ converging to the same limit $p$. Now, either is $||\hat{z}_n||$ finite, which implies that $p\in \mathcal{O}(X)$ and hence a contradiction, or  $||\hat{z}_n||\rightarrow\infty$. Following the same argument as above, $\hat{z}_n/||\hat{z}_n||\rightarrow \hat{w}^i$ where $\hat{w}^i$ is an isotropy. Consequently, we can do the same reduction and reduce the space of null-rotations to $\mathbb{R}^{n-4}$. Eventually, this has to terminate, and we conclude that this case leads to a contradiction.

\noindent \underline{3) Both $||\lambda_n||, ~||z_n||\rightarrow \infty$: } Combining the two cases above we note that we get the action:
\[ (B_{\lambda_n}N[T])_b=e^{b\lambda_n}(N[z^i,T])_b.\] First we select a subsequence so that $z^i_n/ ||z_n||$ converges to a $w^i$. Next, we consider the ``spins''; i.e., the connected subgroup $H\subset K$, which commutes with the boosts $B_\lambda$. This group is isomorphic to $SO(n-2)$ and for $h\in H$, $hB_{\lambda_n}h^{-1}=B_{\lambda_n}$ (it preserves the null-directions $\ell$ and $n$). We can now consider the sequence $h_n$ which aligns $z^i_n$ with $w^i$, so that $h_nN[z^i]h^{-1}_n=N[||z^i||w^i]$. Clearly, since $z^i/||z^i||$ converges to $w^i$, $h_n$ can be chosen so that it will converge to the unit in $H$. By multiplying the group elements $g_n$ with $h_n$ we get $h_ng_n=h_ng_nh_n^{-1}\cdot h_n$, and consequently, $h_ng_n(T)=B_{\lambda_n}N[||z_n||w^i,h_n(T)]$. We note that this sequence converges also to ${p}$ since $h_ng_n(T)=h_n(g_n(T))$. 

It is now advantageous to switch to the dynamical systems point of view which will actually provide with a more general proof (including 1) and 2) above). Using Prop. \ref{flowprop}, we know there exists a point $p$ on the boundary which has a flow so that $e^{\tau{\mathcal X}}(T)\rightarrow p$. From the discussion above, we can assume that: 
\beq
{\mathcal X}=r{\mathcal X}_b+s^i{\mathcal X}_i, 
\label{eq:X}\eeq
where ${\mathcal X}_b$ and ${\mathcal X}_i$ is a Lie algebra basis of boosts and null rotations, respectively, and $r$ and $s^i$ are constants. Indeed, since the above ${\mathcal X}$ is always conjugate to either a pure null-rotation or a boost (see appendix), the proofs of 1) and 2) implies that the theorem is true for case 3) also. 

\begin{figure}[t]
\centering
{\includegraphics[width=60mm]{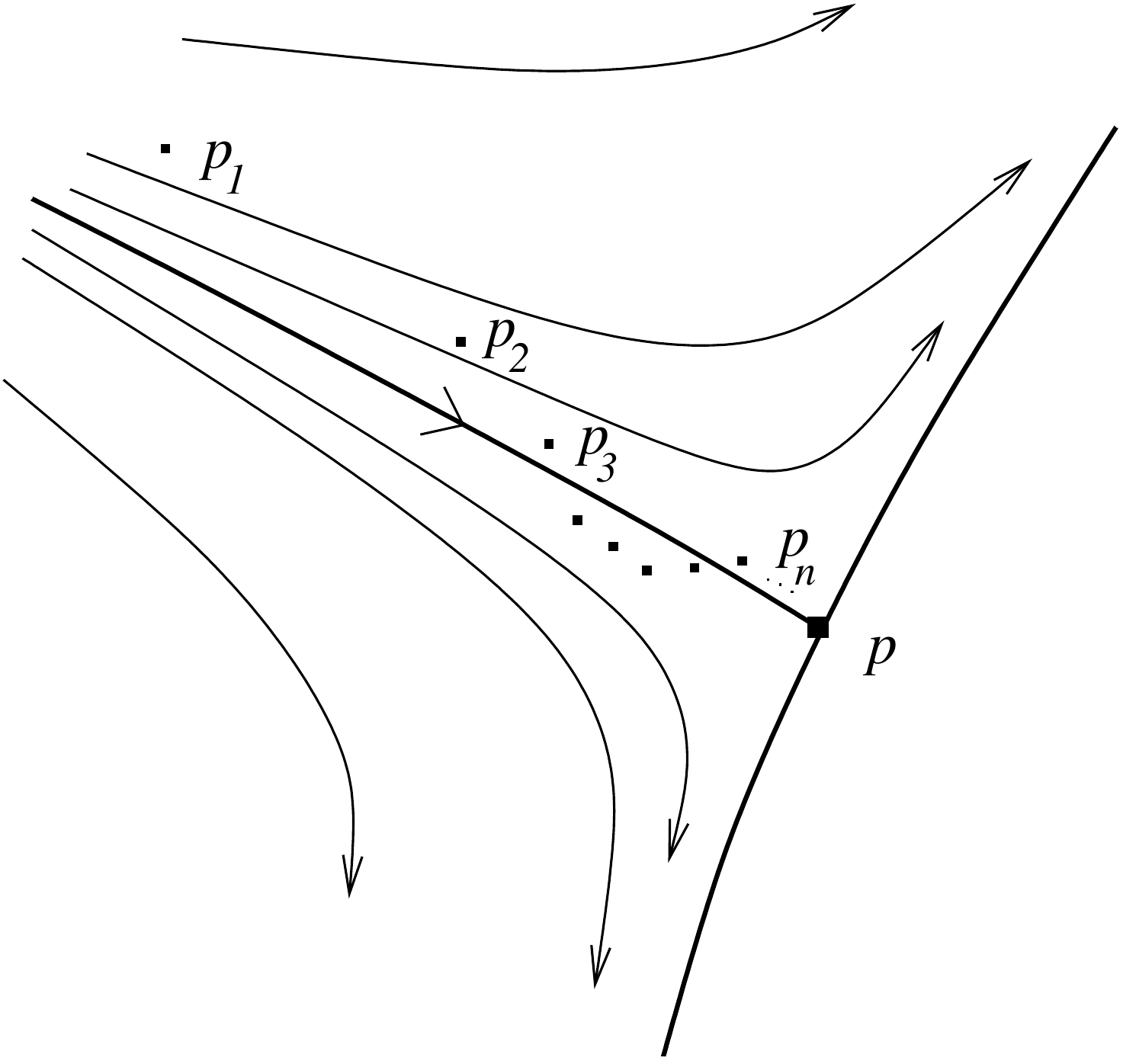}}
\caption{The dynamical system defines a flow for which one solution curve has the point $p$ on the boundary as a limit.}\label{fig:dynsys}
\end{figure}

Figure \ref{fig:dynsys} illustrates the the idea that the dynamical system defines a slow on the orbit $\mathcal{O}(X)$. One solution curve has $p$ as a future limit. In the figure we have also included the sequence $p_n$ as comparison. 

Using eq.(\ref{eq:X}) we note that the boost weight decomposition with respect to  ${\mathcal X}_b$ gives
\[ (B_{\lambda}N[T])_b=e^{b r\tau}(N[\tau s^i,T])_b\rightarrow (p)_b.\]

The polynomial $(N[z^i,T])_b$, where $z^i=\tau s^i$ is now a polynomial in $\tau$ and will diverge as $||z||^{q_b}$, for a non-negative $q_b$, unless $(N[z^i,T])_b=0$. The power $q_b$ is zero for all $b$ if and only iff the $s^i$ is an isotropy. If  the  $s^i$ is an isotropy, we reduce it the same way as in 2). The boost-weight $b$ components $(B_{\lambda}N[T])_b$ will thus be of the order $\sim e^{br\tau} ||z||^{q_b}$ (unless  $(N[s^i,T])_b=0$). Since the point $p$ is finite we thus get the possibilities for each $b$: 
\begin{enumerate}
\item[i)] $q_b>0$, $br<0$.
\item[ii)] $q_b=0$, $br\leq 0$.
 \item[iii)]  $br>0$, $(N[z^i,T])_b=0$.
\end{enumerate}
In the latter case we note that we can choose the null-rotation given by $s^i$ and thus transform the appropriate boost weight components to zero. 

We now see that, in particular, we get the condition that either $\lambda_n\rightarrow \infty$ and $(T)_{b>0}=0$, or $\lambda_n\rightarrow -\infty$ and $(T)_{b<0}=0$. 

Therefore, since the boost acts on all of tensors in $X$, we see that if we apply this to $X=\sum_{b=-N}^N(X)_b$, we end up with the conclusion that either \[ X=\sum_{b=0}^N(X)_b, \quad \text{or} \quad X=\sum_{b=-N}^0(X)_b.\] 
These two statements are equivalent since  the symmetry, eq.(\ref{dsym}), simultaneously swaps all the boost-weights: $(X)_b\rightleftharpoons (X)_{-b}$, for all $(X)_b$. 

This completes the proof.

\paragraph{Remark:} We note that the argument of the proof goes along the following lines: existence of a $p$ on the boundary $\Rightarrow$ existence of a sequence $p_n$   $\Rightarrow$ existence of a sequence $g_n\in G$  $\Rightarrow$ existence of a frame for which the components have non-positive boost weights. 

It thus seems that the existence of $p$ is the key to understand why such an aligned frame exists. Indeed, we note that a $p$ on the boundary but not in the orbit is a very restrictive requirement. For example, consider the simple case where we have a sole vector $v_a$. The case when the vector is not characterised by its invariants is when $v_av^a=0$; i.e., the null-cone, and the only possible $p$ is  the origin: $p=0$. This is indeed a very special case and the orbits not characterised by their invariants are those $v_a\neq 0$, where $v_av^a= 0$.

A non-trivial example is the Ricci tensor, $R_{ab}$. The Segre types not characterised by their invariants are (in 4D) $\{211\}$ and $\{31\}$. In these cases the $p$ needs to be of Segre types $\{(1,1)11\}$ or $\{(1,11)1\}$, respectively. Defining $X_0$ to be the canonical form representing  $\{(1,1)11\}$ or $\{(1,11)1\}$ (i.e., diagonal), then $p$ can be of any form $p=g(X_0)$, for a $g\in O(3,1)$. Thus there exists a $g_n\in O(3,1)$ such that $g_n(X)\rightarrow p$. In this case, it is advantageous to define a new sequence $\tilde{g}_n\in O(3,1)$, where $\tilde{g}_n=g^{-1}g_n$ so that $\tilde{g}_n(X)\rightarrow \tilde{p}=X_0$. Thus this sequence converges to the canonical form and  thus the frame defined by   $\tilde{g}_n$ must approach the canonical form of $X_0$. However, since $||g_n||$ is unbounded, while $||g_n(X)||$ is bounded,  the group sequence $g_n$ needs to attain a very special form. Clearly, the term ``existence of'' in each of the steps in the proof is actually very restrictive and forces the tensors to be of aligned type II. 

\section{Corollaries of the alignment theorem}
We note that the alignment theorem results in a cascade of corollaries, many of which prove previous conjectures. Let us mention some of these. 

All the following results hold for $n$-dimensional Lorentzian spacetimes. 

\subsection{Type D tensors}
In the special case of a type D tensors, we have: 
\begin{cor}
A tensor of type D is characterised by its invariants. 
\end{cor}
\begin{proof}
For a type D tensor we have the boost-weight decomposition $T=(T)_0$ and hence boost-invariant. Assume that the orbit $\mathcal{O}(T)$ is not closed. Then by the proof of the alignment theorem there is a sequence $p_n$ in the orbit converging to a point $p$ on the boundary, but not in the orbit; i.e., $p\notin\mathcal{O}(T)$.  However, since a type D tensor is boost invariant, the argument in the proof implies that $B_\lambda[T]=T$, implying that $p\in\mathcal{O}(T)$ leading to a contradiction. Therefore, $\mathcal{O}(T)$ is closed and the corollary follows.
\end{proof}
This is an interesting corollary and implies that, for example, a CSI spacetime for which all its polynomial curvature invariants are constants and with all its curvature tensors of type D is locally homogeneous. 

Note that spacetimes being og type D to all orders where studied in \cite{invHigher} where it was found that these are all of Kundt type \cite{Kundt}. 

\subsection{Tensors with vanishing invariants}
Consider a tensor $T$ with all vanishing polynomial invariants. Then if the tensor is characterised by its invariants $T=0$. If the tensor is not characterised by its invariants, it is of type II and consequently has no positive boost-weight components:
\[ T=(T)_0+(T)_{-1}+...\] 
Furthermore, we note that (by the previous corollary) that $(T)_0$ is characterised by its invariants and must therefore be zero: $(T)_0=0$. Therefore: 
\begin{cor}
Consider a tensor with all vanishing polynomial invariants. Then there is a null frame such that the tensor has only negative boost weights: 
\[ T=\sum_{b<0}(T)_b.\]
\end{cor}
This is earlier referred to the algebraic VSI conjecture \cite{Higher} and has previously not been proven. However, here we see it follows from the alignment theorem and the previous corollary. Moreover, it proves the important Lemma 10 (whose proof is apparently missing) in the higher-dimensional VSI paper. 
\subsection{CSI spacetimes}
For spacetimes with constant polynomial curvature invariants the alignment theorem provides a proof of the CSI$_F$ conjecture, stated in \cite{CSI}, in higher-dimensions: 
\begin{cor}
For a  CSI spacetime there exists a (null) frame such that either:
\begin{enumerate}
\item all the components of the curvature tensors, to all orders are constants; or, 
\item all curvature tensors can be written on the form 
\[ R=(R)_0+(R)_{-1}+...\] 
where the boost-weight $0$ components are all constants.  
\end{enumerate}
\end{cor}
\begin{proof}
If the metric is characterised by its invariants, then we can determine (at least in principle) the components of the curvature tensors from the invariants. Since this is a CSI spacetime, the invariants are constant over a neighbourhood; consequently, the components of the curvature tensors can be chosen to be constants too.  

If the metric is not characterised by its invariants then the alignment theorem implies that all the curvature tensors are of type II. By considering the boost-weight components $(R)_0$, then the set of polynomial invariants  defined by the boost-weight components $(R)_0$ are the same those from  $R$. Since $(R)_0$ is of type D, these are characterised by its invariants and consequently can be chosen to have constant components. We note also that this frame is consistent with the frame from the alignment theorem.  
\end{proof}
The first of these is the locally homogeneous spacetimes, while the second case is the degenerate case. In 3 and 4 dimensions, it has been shown that the latter case consists of the Kundt spacetimes \cite{3DCSI,4DCSI}. In higher dimensions this is likely to be true, but to date a proof is still lacking. However, a major step on the way to proving this is the above corollary which proves the so-called CSI$_F$ conjecture. 
\subsection{Spacetimes of type I/G}
A converse of the alignment theorem is also worth noting:
\begin{cor}
A spacetime of proper Weyl (or Ricci) type I/G is characterised by its invariants. Indeed, if any of the $k$th derivative Riemann tensors $\nabla^{(k)}R$ is of proper type I or more general, then the spacetime is characterised by its invariants. 
\end{cor}
This was proven in 4 dimensions in \cite{inv}, and in higher dimensions, only parts of this result have been proven. This result is thus a big improvement and it gives a clear-cut separation between the metrics that are characterised by their invariants, and those that are not.

\section{Discussion} 
\subsection{Significance of metrics being characterised by their invariants}
So far only a formal mathematical definition is given as to when a metric is characterised in terms of its invariants. More intutiviely, one can say that \emph{all the local information of the metric is contained in the polynomial invariants.} If we define $\mathcal{I}$ to be the set of all \emph{polynomial curvature invariants}, then this means that local information (except possible some discrete transformations), for example, the existence of Killing vectors are incorporated in the set of invariants, $\mathcal{I}$.  Let us consider the case of Killing vectors for which we have the following result \cite{operator}:
\begin{cor}  
If a spacetime $(\mathcal{M},g_{\mu\nu})$ is characterised by its invariants (weakly or strongly), then if there exists  
a vector field, ${\mbold\xi}$, such that 
\[ {\mbold\xi}(I_i)=0. \quad \forall I_i\in \mathcal{I},\]    
then there exists a set, $\mathcal{K}$, of  Killing vector fields such that at any point the vector field ${\mbold\xi}$ coinsides with a Killing vector field $\tilde{\mbold\xi}\in \mathcal{K}$.  
\end{cor}  
Note that we are not saying that  ${\mbold\xi}$ is a Killing vector, only that it implies the existence of a Killing vector $\tilde{\mbold\xi}$. 

We also note that since tensors of type I/G, D, or O are all characterised by their invariants so this result applies to all metrics having  curvature tensors of either of these types. For these spacetimes we can thus check the invariants to check existence of Killing vectors. Note however, that if we are interested in Killing vectors belonging to the \emph{isotropy group}, then these are zero at that point so ${\mbold\xi}(I_i)=0$ trivally. On the other hand, the isotropy would lead to a relation between the components of the curvature tensors implying certain \emph{syzygies} to be fulfilled. These syzygies, which define certain relations between the invariants, where discussed in \cite{DISC}. In order to check for isotropies one thus need to systematically study all possible syzygies of the curvature tensors. These syzygies are closely related to the corresponding curvature operators of the space \cite{operator}. All possible curvature operators must possess the corresponding syzygies before we would be able to conclude the existence of an isotropy.
 
The alignment theorem has also significance for the classification of spacetimes. For example, it gives a criterion for when spacetimes can be classified using the polynomial invariants only. Interestingly, the only cases where this is not the case is when all the curvature tensors are aligned and of type II (or simpler). If the curvature tensors do not fulfill the alignment theorem then the spacetime is classified using its polynomial invariants. This cleary has great significance for the equivalence principle. 

Note also that the alignment theorem provides a more direct proof of the $\epsilon$-property \cite{epsilon2,epsilon} by directly utilising the boost-weight decomposition. 

\subsection{Outlook}

Degenerate spacetimes are important in many (higher dimensional) theories, for example, topologically massive gravity \cite{TMG}, supersymmetry \cite{super}, holonomy \cite{Holonomy}, and special solutions, like gyratons \cite{gyraton}. Evidently, the alignment theorem proves their special significance in terms of the invariants. Indeed, the alignment theorem gives a clear-cut condition for which spacetimes are degenerate when it comes to their polynomial curvature invariants. The theorem is valid in any dimension and is also valid for any tensor. The alignment theorem thus has applications beyond relativity as any tensor not characterised by its invariants in an Lorentz-signature background would be subject to the alignment theorem. 

The proof of the theorem itself is based on simple real analysis (in particular, convergent sequences) and group theory (through the Iwasava decomposition). This sheds new light on the relation between the tensors and their polynomial invariants. Indeed, these techniques are fairly general and are applicable to other signatures as well. This will be the subject of futher work. 

Although the alignment theorem enabled us to prove a series of previously unproven conjectures and problems there are still a few unanswered questions. For example, are all degenerate metrics Kundt metrics \cite{Kundt}? Clearly, without the alignment theorem, this would be a hard result to prove, indeed, in 4 dimensions this was proven in a case-to-case basis \cite{inv}. However, in higher dimensions this question has been elusive but by using the alignment theorem this may now be tractable. 

\section*{Acknowledgments} 
I would like to thank Robert Milson whose comment on the alignment of tensors triggered me to explore this topic and write this paper. In addition, I would like to thank him and Lode Wylleman for comments on an earlier version of this paper. 

\appendix
\section{Conjugacy classes of $BN$} 

Consider the case of a boost and a null-rotation, $Y=BN$. In a block matrix representation we will write: 
\beq
Y=\begin{bmatrix} 
e^\lambda & 0 & 0 \\
-z^i & 1 & 0 \\ 
-\frac 12||z||^2e^{-\lambda} & z^ie^{-\lambda} & e^{-\lambda} 
\end{bmatrix}, \quad X=\begin{bmatrix} 
1 & 0 & 0 \\
-w^i & 1 & 0 \\ 
-\frac 12||w||^2 & w^i & 1
\end{bmatrix}.
\eeq 
We note that $Y$ is null-rotation and a boost, while $X$ is a pure null-rotation. 

By computing the adjoint action of $X$ on $Y$: 
\beq
{\rm Ad}_{X}(Y)=XYX^{-1}, 
\eeq
we note that we get another boost and null-rotation with respect to $w^i(e^\lambda-1)+z^i$. Therefore, if $\lambda\neq 0$, by choosing $w^i=z^i/(e^{\lambda}-1)$, we get: 
\beq
{\rm Ad}_{X}(Y)=XYX^{-1}=\begin{bmatrix} 
e^{\lambda} & 0 & 0 \\
0 & 1 & 0 \\ 
0 & 0 & e^{-\lambda}
\end{bmatrix}, \quad \text{for} \quad w^i=\frac{z^i}{e^{\lambda}-1}.
\eeq
Therefore, a non-zero boost and a null-rotation is always conjugate (using a null-rotation) with a pure boost.

\end{document}